\documentclass[twocolumn]{aastex63}
\usepackage[normalem]{ulem}   
\usepackage{hyperref}
\usepackage{amsmath}
\usepackage{amssymb}
\usepackage{mathrsfs}
\usepackage{animate}
\usepackage{graphicx}
\usepackage{rotating}
\usepackage{bbm} 
\usepackage{amsthm}        
\usepackage{comment}

\usepackage{svg}

\usepackage{subfigure}
\usepackage{tikz}
\usepackage{mdframed}

\usepackage{lineno}

\newcommand{\ce}{c_\mathrm{e}}
\newcommand{\cp}{c_\mathrm{p}}
\newcommand{\cs}{c_\mathrm{s}}
\newcommand{\Cn}{C_\mathrm{n}}
\newcommand{\game}{\gamma_\mathrm{e}}
\newcommand{\gamp}{\gamma_\mathrm{p}}
\newcommand{\gams}{\gamma_\mathrm{s}}
\newcommand{\kB}{k_\mathrm{B}}

\renewcommand{\mp}{m_\mathrm{p}}
\renewcommand{\ne}{n_\mathrm{e}}
\newcommand{\np}{n_\mathrm{p}}

\newcommand{\Pe}{P_\mathrm{e}}
\newcommand{\Pp}{P_\mathrm{p}}
\newcommand{\Ps}{P_\mathrm{s}}
\newcommand{\rc}{r_\mathrm{c}}
\newcommand{\Te}{T_\mathrm{e}}
\newcommand{\Teo}{T_\mathrm{e0}}
\newcommand{\Tp}{T_\mathrm{p}}
\newcommand{\Tpo}{T_\mathrm{p0}}
\newcommand{\Ts}{T_\mathrm{s}}
\newcommand{\Tso}{T_\mathrm{s0}} 
\newcommand{\uc}{u_\mathrm{c}}

\newcommand{\xs}{x_\mathrm{s}}
\newcommand{\nisos}{n_\mathrm{iso|s}}   
\newcommand{\Psisos}{P_\mathrm{s,iso|s}}
\newcommand{\Tsisos}{T_\mathrm{s,iso|s}}
\newcommand{\risoe}{r_\mathrm{iso|e}}
\newcommand{\risop}{r_\mathrm{iso|p}}
\newcommand{\risos}{r_\mathrm{iso|s}}

\newcommand{\ntils}{\Tilde{n}_\mathrm{s}} 
\newcommand{\sums}{\sum_\mathrm{s = \{p,e\}}}

\newcommand{\p}[1]{{\color{magenta}{#1}}}

\shorttitle{Radial Solar Winds Statistical Analysis and Iso-poly Modeling}
\shortauthors{Dakeyo et al.}
\graphicspath{{./}{figures/}}

\begin{document}

\title{Statistical Analysis of the Radial Evolution of the Solar Winds between 
0.1 and 1 au, and their Semi-empirical Iso-poly Fluid Modeling}

\correspondingauthor{Jean-Baptiste Dakeyo}
\email{jean-baptiste.dakeyo@obspm.fr}

\author[0000-0002-1628-0276]{Jean-Baptiste Dakeyo}
\affiliation{LESIA, Observatoire de Paris, Universit\'e PSL, CNRS, Sorbonne Universit\'e, Universit\'e de Paris, 5 place Jules Janssen, 92195 Meudon, France}
\affiliation{IRAP, Observatoire Midi-Pyrénées, Universit\'e Toulouse III - Paul Sabatier, CNRS, 9 Avenue du Colonel Roche, 31400 Toulouse, France}

\author[0000-0001-6172-5062]{Milan Maksimovic}
\affiliation{LESIA, Observatoire de Paris, Universit\'e PSL, CNRS, Sorbonne Universit\'e, Universit\'e de Paris, 5 place Jules Janssen, 92195 Meudon, France}

\author[0000-0001-8215-6532]{Pascal D\'emoulin}
\affiliation{LESIA, Observatoire de Paris, Universit\'e PSL, CNRS, Sorbonne Universit\'e, Universit\'e de Paris, 5 place Jules Janssen, 92195 Meudon, France}
\affiliation{Laboratoire Cogitamus, 75005 Paris, France}

\author[0000-0001-5258-6128]{Jasper Halekas}
\affiliation{Department of Physics and Astronomy, University of Iowa, Iowa City, IA 52242, USA}

\author[0000-0002-7728-0085]{Michael L. Stevens}
\affiliation{Smithsonian Astrophysical Observatory, Cambridge, MA, USA}

\begin{abstract}

Statistical classification of the Helios solar wind observations into several populations sorted by bulk speed has revealed an outward acceleration of the wind. The faster the wind is, the smaller is this acceleration in the 0.3 – 1 au radial range \citep{Maksimovic2020anticorrelation}. In this article we show that recent measurements from the Parker Solar Probe (PSP) are compatible with an extension closer to the Sun of the latter Helios classification. 
For instance the well established bulk speed/proton temperature $(u, \Tp)$ correlation and bulk speed/electron temperature $(u, \Te)$ anti-correlation, together with the acceleration of the slowest winds, are verified in PSP data. We also model the combined PSP $\&$ Helios data, using empirical Parker-like models for which the solar wind undergoes an "iso-poly" expansion: isothermal in the corona, then polytropic at distances larger than the sonic point radius. The polytropic indices are derived from the observed temperature and density gradients. 
Our modelling reveals that the electron thermal pressure has a major contribution in the
acceleration process of slow and intermediate winds (in the range of 300-500 km/s at 1 au), over a broad range of distances and that the global (electron and protons) thermal energy, alone, is able to explain the acceleration profiles.
Moreover, we show that the very slow solar wind requires in addition to the observed pressure gradients, another source of acceleration. 

\end{abstract}


\section{Introduction} \label{sec:intro}

In the hydrodynamic description, the solar wind comes from the thermal expansion of the million Kelvin solar corona which cannot remain in hydrostatic equilibrium around the Sun. 
Indeed,  as firstly establish by \cite{Parker1958dynamics}, the solar wind is the result of the conversion of the coronal thermal energy into directed kinetic energy. This implies the generation of a flow which becomes supersonic at a distance ($\rc$) of a few solar radii from our star.

Many authors have studied the radial evolution of the thermodynamic properties of the solar wind, using the large coverage of heliocentric distances allowed by the Helios missions \citep{Schwartz1983radial, Hellinger2011heating, Hellinger2013proton, Stverak2015electron, Maksimovic2020anticorrelation}.  In an attempt to disentangle the temporal from the radial variations of the solar wind, \cite{Schwartz1983radial} have applied the technique of radial line-ups, where they have studied a single piece of solar wind as seen at two different heliocentric distances. They have observed a radial compression of the flux tube, that can be an illustration of wind interactions (co-rotating interaction regions). In order to study the heating, \cite{Hellinger2011heating} and \cite{Hellinger2013proton} compare, respectively for the slow and fast winds, the heating needed to get the 
observed proton temperature gradients (parallel and perpendicular), to the heating rates deduced from the radial wind speed. Both studies strongly suggest an efficient transfer of thermal energy from the parallel to the perpendicular direction to be in accordance with the proton temperature gradients. Other authors as \cite{Stverak2015electron} have made similar analysis on the electrons, and have shown that the observed empirical radial profiles do not require any external heat source (heat flux and its divergence) to explain the observed electron temperature gradients, for both slow and fast representative solar wind streams.  

More recently, \cite{Maksimovic2020anticorrelation}, inspired by the work of \cite{Totten1995empirical}, have classified the different winds observed by Helios according to their velocity, imposing the same order between velocity populations at all distances. They have shown that the correlation bulk speed/proton temperature ($u$, $\Tp$) and the anti-correlation bulk speed/electron temperature ($u$, $\Te$), first found around 0.7 au, extends until 0.3 au (the closest approach distance of Helios missions). 
In the present work we use the same wind classification technique as \cite{Maksimovic2020anticorrelation} and extend it to PSP data closer to the Sun.

After Parker's seminal work, a great number of authors have proposed semi-empirical fluid models of the solar wind, imposing remote sensing observations as boundary conditions in the corona \citep{Esser1997_flux_tube_expansion, Cranmer1999spectroscopic, Sanchez2016very}. 
These authors have often used more or less ad'hoc sources of energy, in addition to the thermal one, allowing them to reproduce  observations at 1 au. 
Another approach is to develop solar wind models including the observed polytropic indices as deduced from the temperature and density gradients. 
For instance \cite{Cranmer2009empirical} empirically constrain fast wind modeling by the observed proton and electron temperature radial dependencies, using a turbulent hydrodynamic model.

Coronal observations in coronal holes and streamers can provide observational constraints to solar wind models.
For a medium solar wind ( $\sim$ 350 - 500 km/s at 1 au ) the proton coronal temperature is found in the range 1 - 3 MK, and the electron coronal temperature within 0.5 - 1 MK \citep{Cranmer1999spectroscopic, Cranmer2002coronal, 1998_David_elec_coronal_hole}. However concerning the fast wind, which has been well established to come from coronal holes, the hydrogen kinetic temperatures are possibly as large as 4 - 6 MK \citep{kohl1996measurement, Cranmer2002coronal}. Then, with enough collision coupling in the low atmosphere, the proton temperature is also expected to be in this range. Regarding the temperature of electrons in coronal holes, it is well established to be lower than in the streamer belt.

In the present approach we also develop a semi-empirical model. In contrast with previous works which start from the observed coronal constraints, we rather base our model on the interplanetary observations, then we derive the expected coronal values. To do this, we use a Parker polytropic model far from the Sun which includes proton and electron pressure contributions separately.  In order to avoid excessive coronal temperature, we include an isothermal solution closer to the Sun. This defines our "iso-poly" fluid model. The polytropic indices and temperatures for both the protons and electrons in the interplanetary medium are derived from observations of the two missions Helios \citep{1981_ref_helios} 
and Parker Solar Probe \citep{fox2016solar}.  

In Section \ref{sec:obs_helios_psp}, we first describe the data sets we use, and how we define the different wind populations. Then, we analyze how the new PSP data compare to the Helios ones within the overlapping range of solar distances. After that, we classify the PSP data the same way as for Helios, and we check whether the radial trends observed for the bulk speed and the temperature gradients in the 0.3 - 1 au range, could be extended closer to the Sun. 
In Section \ref{sec:obs_model},  we describe our iso-poly fluid model, and the way its free parameters are constrained by the observations. 
Finally, we summarize our results in Section \ref{sec:conclusion}.  More information and details on the iso-poly model are provided in the appendixes \ref{sec:appendix_iso-poly_detailed_calculs} - \ref{sec:appendix_identif_param}.

\begin{figure*}[t]
    \hspace{-0.5cm}
    \includegraphics[scale=0.56]{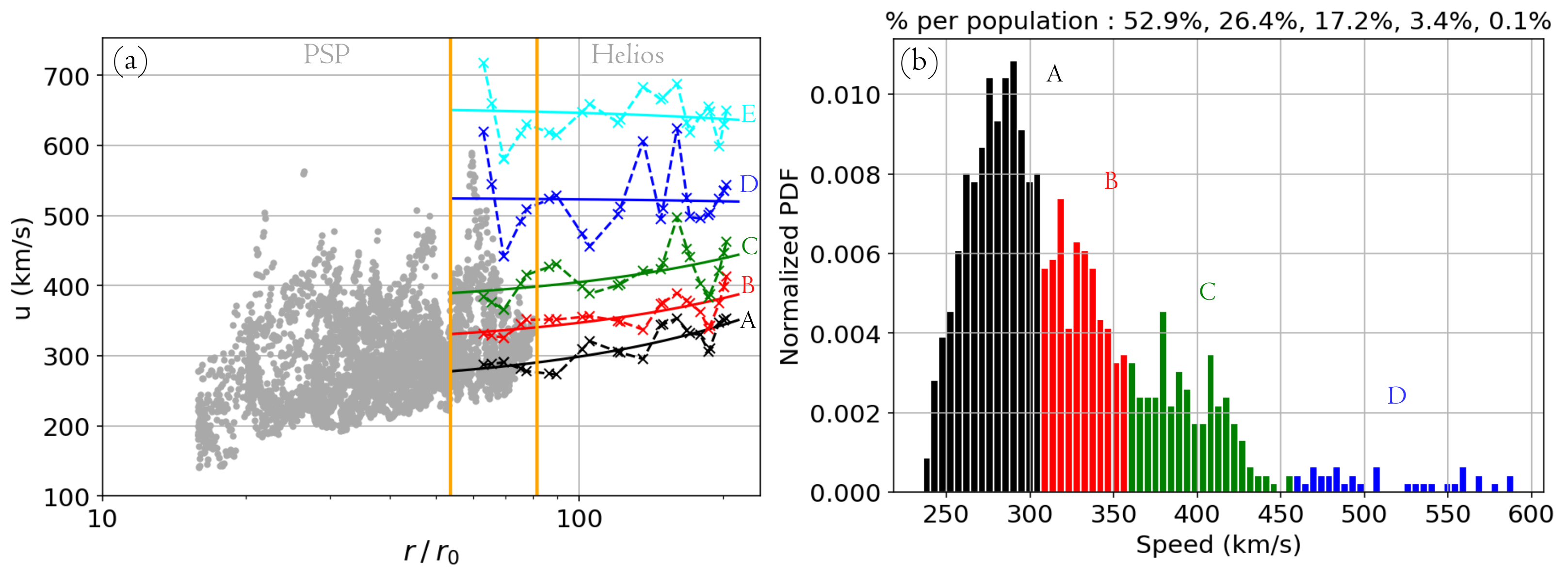}
    \vspace{-0.55cm}
    \caption{ 
(a) Median proton bulk speed, $u(r)$, in colored dashed lines for 5 populations from Helios measurements of bulk speed between $0.3$ and $1$ au. The data are first regrouped within radial bins, then the wind populations are defined with quantiles \citep{Maksimovic2020anticorrelation}. The linear fits of the Helios speed populations, $u(r)$, are shown with colored solid lines.  The PSP measurements from SPAN-Ai and SPC instruments are plotted with grey points. The two orange vertical lines delimit the overlap interval of SPC and Helios data (0.28 - 0.38 au). 
 (b) Probability distribution function of bulk speed of SPC on the overlap interval PSP - Helios. These data are classified using the Helios quantiles. }
    \label{fig_classif_population_Helios}
\end{figure*}

\section{Wind Populations from Helios and PSP Observations} \label{sec:obs_helios_psp}

\subsection{Revisited Helios Measurements}
\label{sec:obs_helios_psp_1}

In this section, we revisit the analysis made by \cite{Maksimovic2020anticorrelation} by removing from the datasets the periods corresponding to interplanetary coronal mass ejections (ICMEs). This was not done in the original study. 
We use two of the Helios data sets used by \cite{Maksimovic2020anticorrelation}. They are derived from the ion and electron electrostatic analyzers on board the Helios 1 and 2 spacecraft \citep{Schwenn1975plasma}. The first data set contains $\sim$ 1 877 000 measurements of proton density $\np$, temperature $\Tp$ and bulk speed $u$. 
The second one, made by \cite{Stverak2009fraction_strahl}, contains $\sim$ 66 000 measurements of electron density $\ne$ and temperature $\Te$. One can find more details about the used Helios data set in \cite{Maksimovic2020anticorrelation}. We also choose to only keep the Helios measurements during the minimal solar activity (from 1974 until 1977), in order to be able to compare the same solar activity level with the PSP observations. 

We remove ICMEs from our Helios data set using the criteria of \cite{Elliott2012temporal}.
We discard the measurements for which at least one of the following criteria on the $\beta$ of the plasma, the proton temperature $\Tp$, and the ratio of the alpha to proton density $n_\alpha /\np$, is satisfied: 
  $(i)$  $\beta < 0.1$, 
  $(ii)$ $n_\alpha /\np  > 0.08 $, 
  $(iii)$ $\Tp / T_{ex} < 0.5$, 
where $T_{ex}$ is a temperature predicted by a scaling law established by \cite{Lopez1986solar} and rescaled with solar distance. In addition to these criteria, we remove for every detected ICME of at least 6 hours long, the 24 hours before and 15 hours after it. Finally, we assume that winds measurement faster than 800 km/s could be possible ICMEs, so that we also remove them.
Our final Helios data set contains $\sim$ 686 000 proton measurements and $\sim$ 65 000 electron measurements.
 
With such data, a possible way to study the solar wind evolution with distance is to classify it into wind populations, determined by a statistical classification of protons speed measurements at different radial distances, as it was done by \cite{Maksimovic2020anticorrelation}. Wind speed observations are first split in radial bins, then for each bin, the bulk velocity distribution is divided with quantiles to classify winds depending on their speed. The median of each speed population 
is kept. This defines a set of median velocities versus distance as shown in dashed colored lines in Figure \ref{fig_classif_population_Helios}a. 
This classification method assumes that the wind population order does not change with solar distance.  
We have made the same choice as \cite{Maksimovic2020anticorrelation} to set 5 wind populations, named from \textbf{A} for the slowest one, to \textbf{E} for the fastest one. This choice of the number of populations is somewhat arbitrary, but we have verified that the results of our study do not depend on this number. The Helios populations have wind speeds ranging between 250 km/s and 700 km/s (Figure \ref{fig_classif_population_Helios}a).
The slower the wind is, the more progressive is its acceleration with radial distance, until the \textbf{E} wind for which the speed is approximately constant in the studied range.
Note that our slow wind population is very similar to the "very slow solar wind" studied by \cite{Sanchez2016very}.

\subsection{PSP Measurements}
\label{sec:obs_helios_psp_2}

There are on board PSP three instruments part of the SWEAP suite \citep{Kasper2015SWEAP} which measure solar wind bulk speed, temperature and density: the Solar Probe Cup (SPC), the Solar Probe ANalyser Ion (SPAN-Ai) both for protons, and the Solar Probe ANalyser Electron (SPAN-E) for electrons. 
The purpose of the present subsection is to establish a single PSP data set, associating for each of the individual times of measurements, one proton and one electron measurement over the largest possible range of solar distances.

Since the SPC instrument is based on the classical design of a Faraday Cup, which measures the protons along the radial field of view, its data have some drawbacks close to the Sun. Because the probe has a very large tangential speed close to the Sun, fewer solar wind protons can enter the radial field of the instrument, causing the measurements to be biased. The slower the wind speed is, the more this effect is important, especially around perihelion since the tangential probe speed has the same order as the radial slow wind speed. Looking to the encounters 4 to 9 SPC data on the relevant servers, we have observed empty regions of measurements closer to the Sun, partly due to this effect. We have thus decided to remove SPC data under 0.2 au ($ \sim \: 43 \: R_\Sun$) to avoid these gaps. 

The SPAN-Ai (SPI) instrument is performing more efficiently closer to the Sun than farther away, because of a better configuration of the field of view due to the tangential motion of the spacecraft  \citep{Kasper2015SWEAP}. The SPI Data Release Notes from NASA documentation indicates that the instrument mainly provides data below $0.25 $ au ($ \sim \: 53 \: R_\Sun$). Therefore, we need SPC data at larger solar distances to have an overlap of distances with Helios data.

With the aim of making the radial coverage between SPI and SPC data instruments, we have compared the speed and the temperature (L3 moments) given by the two instruments. 
The speeds are comparable, while the temperatures are not as close. 
Indeed, comparing only the the radial temperature moment for the two instruments during periods where the solar wind proton peak falls in the join field of view for both SPAN-Ai and SPC, we note that these measurements typically differ by $T_{r|SPI} \sim 2 \: T_{r|SPC}$. Secular trends in time and space are consistent between the two instruments, suggesting that the difference between the two must be some systematic error. An inspection of proton core peak widths over such periods shows consistency between the two instruments, however the SPAN-Ai instrument consistently resolves the extended tails of the proton distribution function out to more extreme speeds and lower fluxes (**Davin Larson, Michael Stevens, private communication**). We therefore hypothesize that the systematic error is a manifestation of the energy partition between the Maxwellian or nearly-Maxwellian part of the proton core and the remaining non-thermal part of the solar wind proton distribution function, where the SPC measurement is dominated by the former while SPAN-Ai moment includes the latter.

To generate a consistent temperature record that combines both SPAN-Ai and SPC in order to cover the largest range of solar distances, we have applied an empirical factor of 2 to the SPC temperatures that is designed to incorporate the non-thermal tail component. We have furthermore empirically adjusted the SPC measurements to account for proton anisotropy, as the SPC measurement is purely radial. For that correction, we use the ratio $T_{r|SPI}/T_{tot|SPI}$ which evolves approximately linearly with solar distance, providing a linear anisotropy ratio with radial distance (Appendix A).

Doing so an equivalent total proton temperature is established from $T_{r|SPC}$ assuming the same anisotropy ratio over the distances covered by SPC. In this way we set an equivalent 3D total proton temperature on larger solar distances.

The SPAN-E (SPE) instrument measures the full electron VDFs in the solar wind. The electron data we use are obtained with the fitting techniques described in 
\cite{Halekas2020electrons}. The total temperature and total density have been obtained by integration of the VDFs after removing the photo-electron and secondary electron contribution.

Considering all the experimental limitations, we have used SPI data below 0.25 au, SPE data below 0.37 au and SPC data from 0.2 to 0.37 au.  For every measurement time where we have both SPI and SPC data, we have kept the mean value. For density measurements, we have made the choice to show only $\ne$ data from SPE \citep{Halekas2020electrons}. Indeed, without measurements of the alpha particles density on the entire studied radial range, it is more relevant to use $\ne$ to estimate the total density of the plasma. 

The PSP observations cover 5 encounters, from E4 to E9, combining in total 2237 hours of measurement for $u$, $\Tp$, $\Te$ and $\ne$. The quantity of data to treat are large because of the high sample rates, especially for SPC. Since in any case we bin the data by distance, we have computed average values of the above parameters over 30 minutes. 

The bulk speed averages are shown as grey dots in Figure \ref{fig_classif_population_Helios}a. It appears that PSP has mainly measured slow and intermediate winds (from 150 km/s to $\sim$ 500 km/s) since its launch. Indeed, since the observational interval corresponds to a period of minimum of solar activity, the fast solar wind in the ecliptic plane is rarely measured.

\begin{figure*}[t]
\hspace{-0.5cm}
\includegraphics[scale=0.72]{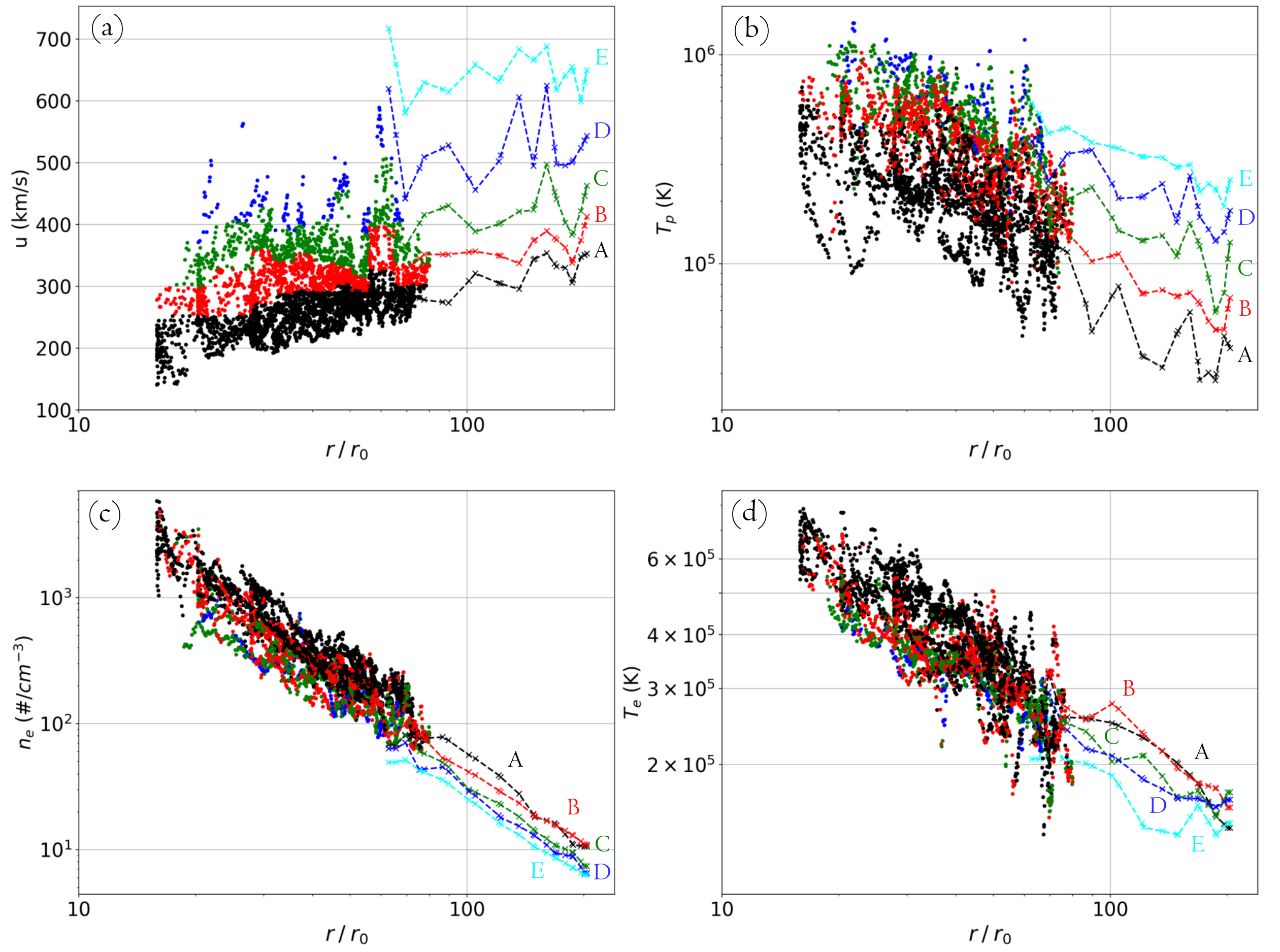}
\vspace{-0.4cm}
    \caption{ Classification of $u(r)$, $\Tp(r)$, $\Te(r)$ and $\ne(r)$ from PSP data set (colored dots). The PSP population percentage is defined in the overlap interval (Figure \ref{fig_classif_population_Helios}) by assigning each PSP data to the closest Helios population. 
    The median values of the 5 Helios populations are added with colored dashed lines as in Figure \ref{fig_classif_population_Helios}a.   
    }
    \label{fig_nuage_pts_Helios_PSP}
\end{figure*}

\begin{figure*}[t]
    \hspace{-0.5cm}
    \includegraphics[scale=0.72]{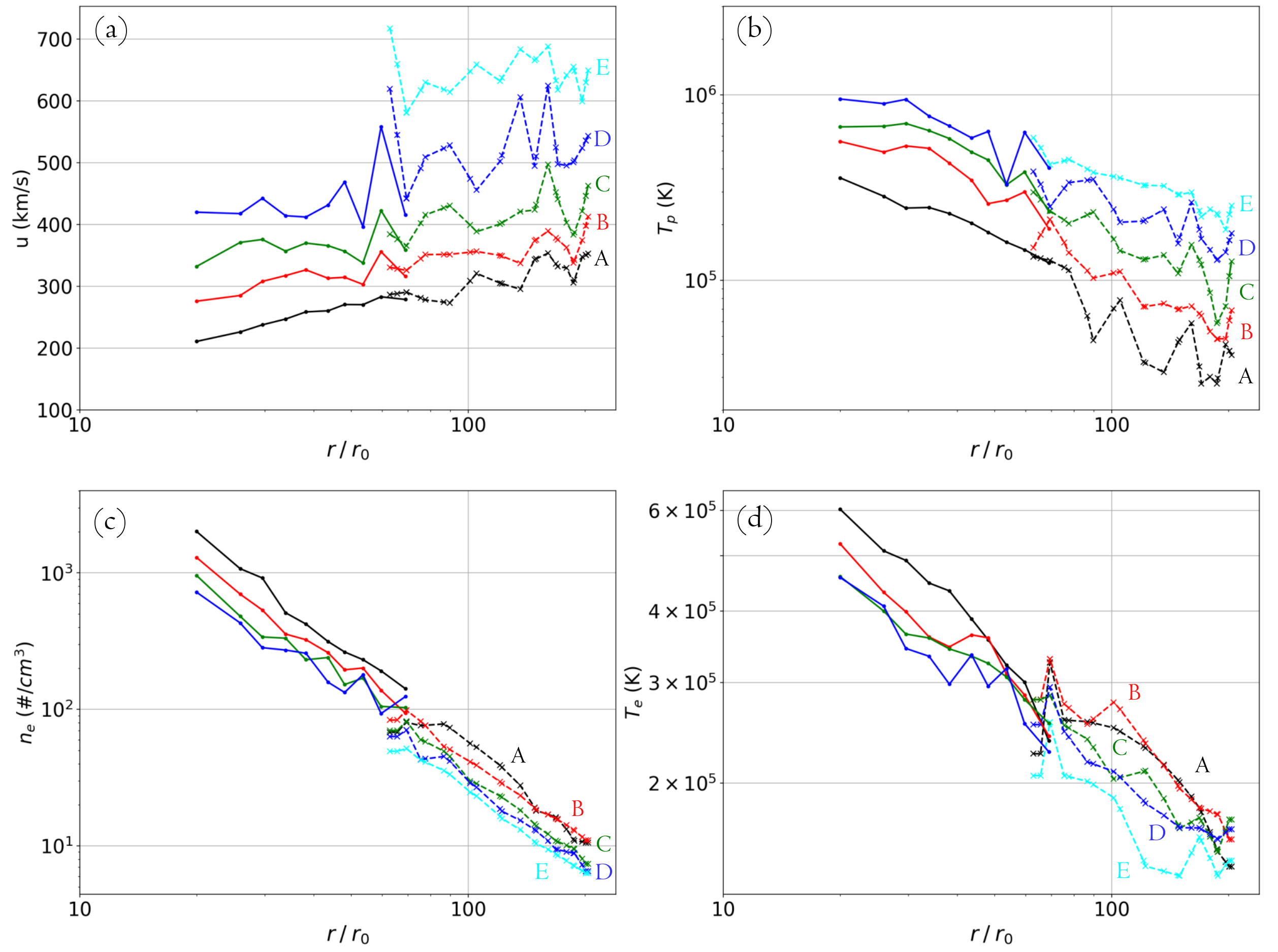}
    \vspace{-0.2cm}
    \caption{ 
    Classification of $u(r)$, $\Tp(r)$, $\Te(r)$ and $\ne(r)$ from PSP data set using the percentages defined with Helios data in the overlapping distances of PSP and Helios. The median values of PSP radial bins are represented with points connected with solid lines. The median values of Helios radial bins are represented with crosses connected with dashed lines.
    }
    \label{fig_med_pts_Helios_PSP}
\end{figure*}

\subsection{How well PSP Data Extend Helios ones?} 
\label{sec:obs_helios_psp_3}

With the purpose of defining the wind populations closer to the Sun, we have determined to what extent the Helios populations are represented in the PSP data coverage. To do this, we have defined an overlap interval between Helios and PSP, represented by the two orange vertical lines in Figure \ref{fig_classif_population_Helios}a. This overlap ranges from 0.28 to 0.38 au. 

In the overlap interval, we have classified the PSP data points (grey points on Figure \ref{fig_classif_population_Helios}a) attributing them one of the populations defined from Helios measurements. For this purpose we compare the bulk speeds between the two probes. We assign each PSP measurement to the Helios population to which it is the closest. To have smoother representation of the Helios median populations profiles (colored dashed lines on Figure \ref{fig_classif_population_Helios}a), we have considered their main tendency using the least square fitted straight lines (represented in solid lines on the same panel). 

The percentage of PSP data corresponding to the Helios populations \textbf{A} to \textbf{E} are represented on Figure \ref{fig_classif_population_Helios}b with the same color code as for the previously established populations. It appears that the two fastest wind populations are much less represented than the others in PSP measurements, with only 3.4 $\%$ for the wind \textbf{D} and 0.1 $\%$ for the wind \textbf{E} within the overlap interval. Therefore, the wind \textbf{E} cannot be studied close to the Sun.

The determined percentages ensure that the PSP measurements are classified in accordance with Helios populations. Next, we divide PSP radial range in 10 intervals with the same number of data, then we apply the quantile classification with the percentage obtained from the overlap interval. Instead of quintiles (20$\%$) for each of the HELIOS populations \textbf{A} to \textbf{E}, we classify the PSP data, within a given radial bin, according to the percentages defined in Figure \ref{fig_classif_population_Helios}b (52.9$\%$ for the wind A, 26.4$\%$ for wind B, 17.2$\%$ for C and the remaining 3.5$\%$ for D). The PSP data with this classification are shown on Figure \ref{fig_nuage_pts_Helios_PSP}. 

We observe a continuation of the radial trends 
between Helios and PSP, for all the displayed parameters (while only the bulk speed is used in the overlap interval to define the PSP populations).
Also some regions without bulk speed and proton temperature data, particularly close to the largest and for the smallest solar distances, are visible on Figure 2. 
Indeed, PSP has not spent enough time to measure each population at all solar distances with enough statistics.
So for populations which are close to these regions, the analysis might be taken with caution. 

Then, we compute for each radial bin the median value for all the populations and for all quantities. These are displayed by dots connected with solid lines in Figure \ref{fig_med_pts_Helios_PSP}. 
As previously mentioned for Figure \ref{fig_nuage_pts_Helios_PSP}, the PSP populations are globally the continuation of the Helios ones. 
This is true for the amplitude of the bulk speed following the definition of the populations in the overlap interval.  However, the speed trends are also comparable between PSP and Helios, while 
constrained.  

Next, proton temperatures from PSP are consistent with a continuation of Helios data (with large fluctuations for the wind \textbf{D} in the overlap interval due to data gaps). Electron densities of PSP data extend the Helios power laws closer to the Sun. Finally, electron temperatures have large fluctuations in and around the overlap interval.  Still outside this region, PSP data are in accordance with the extensions of the Helios power laws with  distances.

The higher variations for the farther radial point of the wind \textbf{B}, \textbf{C} and \textbf{D} for all PSP parameters are probably due to the regions empty of data in Figure \ref{fig_nuage_pts_Helios_PSP} as mentioned above.
Next, we notice that the proton temperature for the winds \textbf{B}, \textbf{C} and \textbf{D}, seems to stop increasing closer to the Sun.  However, considering the lack of good statistical coverage for the radial bins closer to the Sun, no definitive interpretation can be made presently.
Indeed, to have a more reliable analysis of the slope in the radial bins closer to the Sun, a longer observational time interval is necessary.

All the observational results shown on the PSP radial range regarding the radial dependencies of the temperature populations and the acceleration of the slow solar wind, are also confirmed by a different radial analysis method of the solar winds evolution of \cite{jasper_2022}.

\section{Comparison between Observations and Modeling Results} \label{sec:obs_model}

\subsection{Hydro-dynamic Model Limitations}
\label{sec:obs_model_1}

Isothermal Parker's model solutions are convenient because they can provide both an analytical expression of the sonic point location and the dependence of the terminal bulk speed with coronal temperature. However, this description requires an infinite energy deposit to maintain the isothermal temperature at all distances. 

A more physical description implies taking into account that the observed solar wind temperature is decreasing with distance. As discussed in Section \ref{sec:intro}, solar wind fluid models using observed polytropic indices have already been proposed. 
However extrapolating temperature back to the corona, the deduced proton coronal temperature is too high compared to spectroscopic observations, especially for fast winds. 

Thus, it could be interesting to mix up isothermal and polytropic approaches. 
An isothermal expansion can produce a supersonic wind relatively close to the Sun. At larger distances, a polytropic expansion only mildly accelerates the wind, while it reproduces the observed decrease of temperature with distance. This iso-poly model takes the best of each approach while putting aside their respective major physical issues (infinite deposit of energy for the isothermal case, and too high coronal temperature for the polytropic one). 
Note that \cite{1960_Parker} has also proposed a two thermal part model, with an isothermal evolution close to the Sun, then adiabatic farther away. However, this description disagree with in-situ measurements of temperature (magnitude and radial evolution), and with the observed acceleration of slow winds on large distance.

\subsection{Iso-poly Solar Wind Model}
\label{sec:obs_model_2}

The equations we develop in this article embed the possibility of two consecutive thermal regime for the solar wind.
We set the following hypotheses: 
  $(i)$ We consider a bi-fluid constituted of electrons and protons, with $u_e = u_p = u $, $\ne = \np = n $, $\Te \neq \Tp$ and $\game \neq \gamp$, with no electric current. 
 $(ii)$ We take into account the thermal pressure gradients and gravity as a source of external force.
   $(iii)$ The problem is studied in the hypothesis of spherical symmetry: $ \quad \partial / \partial \theta =  \partial / \partial \phi  = 0 $. 
$(iv)$ A stationary flow is modeled.
   $(v)$ The non-thermal effects of the magnetic field on the plasma are neglected. 

The transition between the two thermal regimes, isothermal and polytropic, is set at the radius $\risop$ and $\risoe$ respectively for protons and electrons. These distances are expected to be different for these two species because different heating/cooling processes are present and because a low collisional coupling occurs between the two species in the considered radial range \citep{Cranmer2002coronal}. 
Next, in order to simplify the expressions below, we only specify the species with s = \{p, e\}, and we write the sum over these species when needed. 

The model incorporates in-situ observational constraints for both electrons and protons with a polytropic law:
  \begin{align}
    \Ts(r) = \Tso \: \bigg( \frac{n(r)}{\nisos} \bigg)^{\gams -1} 
           = \Tso \: \ntils^{\gams -1}  \,.
    \label{eq:expression_Ts}
  \end{align} 
where $\gams$ is constrained by in-situ observations to be uniform in the PSP and Helios radial range, while dependant of the wind population and specie.  We introduce the density at $r=\risos $, $\nisos$, within Equation \eqref{eq:expression_Ts} in order to have a formula valid both for $r<\risos$ ($\gams=1$, isothermal, $\Ts(r) = \Tso $) and for $r>\risos$ ($\gams > 1$, for constant value), and $\Ts(r)$ is continuous at $r=\risos$.  The notation $\ntils = n(r)/\nisos$ is introduced to simplify the writing of the following equations.

The conservation of momentum is written as:
  \begin{align}
   n\, \mp\, u \frac{d u}{d r} &= - \sums \frac{d \Ps}{d r} 
                                  - n\, \mp \frac{G \, M}{r^2} \,.
  \label{eq_momentum_sans_hypothese}
  \end{align}
For a given thermal profile, e.g., Equation \eqref{eq:expression_Ts}, the pressures $\Ps$ are proportional to the plasma density.  Indeed, all terms in Equation \eqref{eq_momentum_sans_hypothese} are linear in $n$ so multiplying the density by any factor (independent of $r$) has no effect on $u$.
The pressures $\Ps $ are written similarly to temperatures in Equation \eqref{eq:expression_Ts}:
  \begin{align}
     \Ps = \Psisos \: \bigg( \frac{n}{\nisos} \bigg)^{\gams} 
         = \Psisos \: \ntils^{\gams}  . 
         \label{eq:expression_P_s}
  \end{align}
Then, Equation \eqref{eq_momentum_sans_hypothese} is rewritten as:
  \begin{align}
    n \, \mp u \frac{d u}{d r} =& 
     - \sums \bigg( \Psisos \frac{d\ntils^{\gams} }{d r} \bigg) 
     - n \, \mp \frac{G \, M}{r^2} \, .
    \label{eq:momentum_detailed}
  \end{align}

Next, the computation of $n(r)$ is deduced only from $u(r)$ using the mass flux conservation:
  \begin{align}
    n(r) = \frac{C_{n}}{u \, r^2} \,,
  \label{eq:n(r)}
  \end{align} 
where $\Cn$ is a constant to be determined for each population with a fit to the in-situ data.  This Equation allows to eliminate $n$ in Equation \eqref{eq:momentum_detailed}.
After several steps of calculation, Equation \eqref{eq:momentum_detailed} is transformed to outline the critical or sonic point located at $r=\rc$ (see Equation \eqref{eq:appendix_momentum_all_term_detailled} in Appendix \ref{sec:appendix_iso-poly_detailed_calculs}). 
This allows to define the transonic solution for which the derivative $du/dr$ is non-zero for all $r$ values. 
Finally, $u(r)$ is numerically computed (see Appendix \ref{sec:appendix_iso-poly_detailed_calculs}), then $n(r)$ and $\Ts(r)$ are computed with Equations \eqref{eq:n(r)} and \eqref{eq:expression_Ts}.

\begin{table}[t]
\scalebox{1.15}{
\hspace{-1.7cm}
\begin{tabular}{cccccc}
    \hline 
    Wind type & A & B & C & D & E
     \tabularnewline
    \hline 
    \hline 
    $\gamp$ & 1.57 & 1.59 & 1.52 & 1.44 & 1.35
     \tabularnewline
    $\game$ & 1.29 & 1.24 & 1.23 & 1.23 & 1.21
     \tabularnewline
    $\risop$ ($R_\Sun$) & 16.1 & 16.4 & 13.6 & 9.2\; & 2.9\;
    \tabularnewline 
    $\risoe$ ($R_\Sun$) & 15.0 & \;9.8 & \; 10.3 & \; 8.0 & \;3.1 
    \tabularnewline 
    \hline
    $\Tpo$ (MK) & 0.65  & 1.10 & 1.63 & 2.51 & 5.61
     \tabularnewline
    $\Teo$ (MK) & 0.79 & 0.81 & 0.71 & 0.75 & 0.88
     \tabularnewline
    $u_0$ (km/s) &  0.001  & 0.07 & 0.6\; & 7 & 104 
     \tabularnewline
    $u_\mathrm{1 au}$ (km/s) & 292 & 354 & 406 & 488 & 634 
     \tabularnewline
    \hline
\end{tabular}  
}
\caption{Parameters of the iso-poly model (four top lines).  Resulting the coronal temperatures $\Tpo$, $\Teo$, and the coronal and at 1 au velocities $u_0$ and $u_\mathrm{1 au}$ (four bottom lines). The parameters are defined by least square fitting the model to temperatures and velocities derived from PSP and Helios measurements for the wind populations from \textbf{A} to \textbf{D}, and only from Helios measurements for the population \textbf{E} (see Figure \ref{fig_med_pts_Helios_PSP_iso_poly}).}
\label{tab:param_fit}
\end{table}

\begin{figure*}[t]
\hspace{-0.5cm}
\includegraphics[scale=0.72]{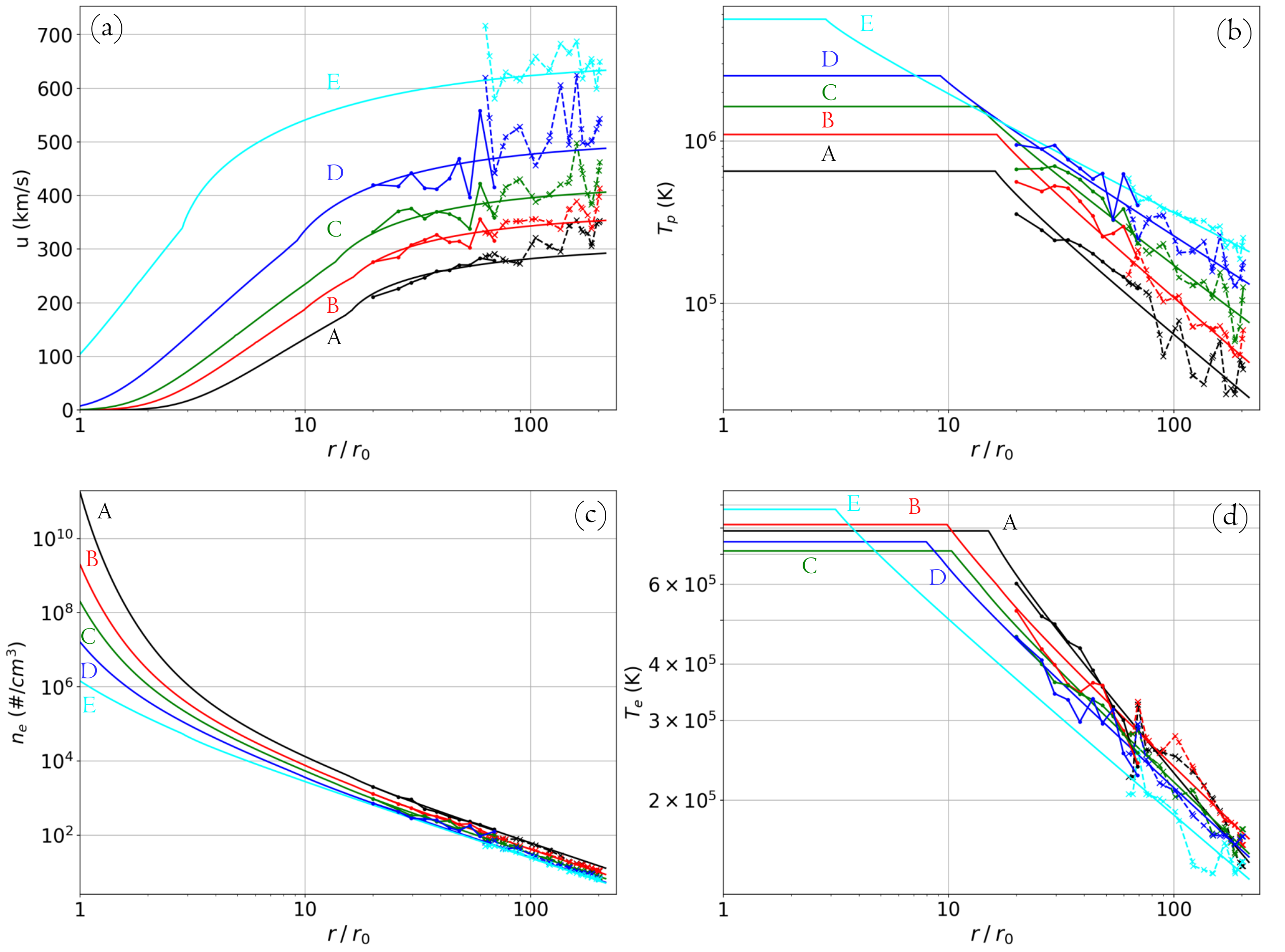}
\vspace{-0.5cm}
    \caption{ Median values of $u(r)$, $\Tp(r)$, $\Te(r)$ and $\ne(r)$ for PSP data (continuous lines linking colored dots) and for Helios data (dashed lines linking colored 
    dots). The iso-poly model curves associated to each family is added with colored continuous lines.  They are obtained by least square fitting the radial data profiles. }
    \label{fig_med_pts_Helios_PSP_iso_poly}
\end{figure*}

\subsection{Iso-poly Modeling of the Wind Populations}
\label{sec:obs_model_3}


We describe below how the iso-poly model parameters are constrained with in-situ data. 
Our iso-poly model has a priory six free parameters: $\gamp$, $\game$, $\Tp$, $\Te$, $\risop$ and $\risoe$.

The polytropic indices can be determined from temperature and density gradient observations. 
Indeed, considering power law evolution of the form $\Ts(r) \: \propto \: r^{\alpha}$, and  $n(r) \: \propto \: r^{\beta}$, the polytropic relation between density and temperature implies $\gamma = (\beta + \alpha)/\beta$.
From the mass flux conservation, density for proton and electron are weakly dependent of $u(r)$ profile  once the main acceleration region is overcome, thus $\beta \approx -2$.  We operate a least square fit on $\Tp(r)$ and $\Te(r)$.
The radial dependence of $\Tp$ and $\Te$ is not the same for all the wind populations, so we fit the $\alpha_p$ and $\alpha_e$ using a linear regression in a log/log space independently for each wind populations. The fitted values of $\gamp$ and $\game$ are summarized in Table \ref{tab:param_fit}.

The fitted power-law of $\Ts(r)$ for each speed population implies that when $\risos$ is defined, $\Tso$ is also defined in order to have a continuous temperature. Then, the only two parameters which need to be defined are $\risop$ and $\risoe$.  This is realized by performing a $\chi^2$ minimization between the model and observed velocities (details in Appendix \ref{sec:appendix_identif_param}).
The radial range for the $\chi^2$ minimization is set to $r < 0.5$ au ($\sim 105 \: R_\Sun$) for all the populations. This minimisation is less constrained for faster solar winds, especially the wind \textbf{E} by the lack of data closer to the Sun. 

The iso-poly curves associated to the parameters in Table \ref{tab:param_fit} are plotted in Figure \ref{fig_med_pts_Helios_PSP_iso_poly} with solid lines. As expected the modeling of proton and electron temperatures is globally in accordance with measurements for all the populations. Locally the proton temperature of the modeled profiles \textbf{B}, \textbf{C} and \textbf{D} are overestimated compared to measurements for the closest radial bins to the Sun. However, considering the empty data regions previously discussed in Figure \ref{fig_nuage_pts_Helios_PSP}a, 
measurement profiles for these radial bins are expected to be raised closer to the model curves with larger statistics. 

The derived coronal proton temperature for all the populations, except \textbf{E}, are in the range of observed coronal temperatures in the solar source regions \citep[1 - 3 MK,][]{Cranmer1999spectroscopic}.  Similarly, the derived coronal electron temperatures are also in agreement with the observed range of 0.5 -1 MK \citep{1998_David_elec_coronal_hole, Cranmer2002coronal}. 

All the iso-poly speed profiles globally fit well to the PSP and Helios measurements (Figure \ref{fig_med_pts_Helios_PSP_iso_poly}a). 
Still, the iso-poly velocity of population \textbf{D} is a bit 
underestimated on the Helios radial range. This could be explained by the fact that few fast winds have been observed by PSP, especially close to the Sun (Figure \ref{fig_nuage_pts_Helios_PSP}a). Then, this decreases the iso-poly velocity since the least square fit is limited to $r < 105 \: R_\Sun$.
Such a difference between observed and iso-poly speeds is not present for the lower speed winds \textbf{C} and \textbf{B} where the number of data points is much larger.  However, the model curve for the slowest wind \textbf{A} is lower than its corresponding measurement curve when going farther from the Sun (Figure \ref{fig_med_pts_Helios_PSP_iso_poly}a). This indicates that the observed proton and electron pressures are not efficient enough to accelerate the solar wind as observed. Therefore, the very slow solar wind 
has another source of acceleration which does not heat the plasma. 

The iso-poly model of the fastest wind \textbf{E} incorporates only Helios data. The model fits well to all the observed variables (Figure \ref{fig_med_pts_Helios_PSP_iso_poly}); nevertheless, the modeled proton coronal temperature of 5.6 MK is much higher than the 1 - 3 MK observed in the corona. Notice that it is still in the order of the 4 - 6 MK hydrogen temperature observed by \cite{kohl1996measurement} in coronal holes (possible of the same order as the protons). 
Concerning the bulk speed close to the Sun, its amplitude is much higher than for other winds, reaching more than 100 km/s at 1 $R_\Sun$. This is high for an initial solar wind bulk speed compared to its value at 1 au, however this is not in contradiction with the speed observations made by \cite{Sheeley1997_coronal_V} close to the Sun. Indeed, they have observed at 2 $R_\Sun$ that wind bulk speed can reach 200 $\sim$ 250 km/s. It concerns mainly slow winds (< 400 km/s at 1 au) but the order of magnitude indicates the possibility of large coronal bulk speeds.
This means that the acceleration provided by the observed proton and electron pressure gradients could be not efficient enough to accelerate the fastest wind (in the hypothesis of coronal temperatures in order of 1 - 3 MK). Consequently, the fast wind implies close to the Sun, below 0.3 au (65 $R_\Sun$), a temperature higher that the typically observed one in the low corona, and/or another source of energy which accelerates the plasma and does not heat it.

The median plasma densities are well ordered within PSP distance range. The density is anti-correlated with the wind speed as observed by Helios and Ulysses \citep{Marsch1989_Te,Gloeckler2003_anticorrelation_u_Te}.  At lower solar distances, the densities no longer follow a power law because of the acceleration of $u(r)$ (Equation \eqref{eq:n(r)}). The wind densities are predicted by the iso-poly model to spread over a much larger range, up to 5 decades, when getting closer to the Sun. 
Next, we compare these densities to the ones derived during a solar cycle minima (1996) from LASCO coronagraph. Even if the studied in-situ data are taken close the ecliptic, we compare the fast wind density to the one observed around the solar poles, which are known to be the source of mostly fast solar wind. The densities derived from the iso-poly models are compatible with the measurements made above 2 $R_\Sun$ \citep[][and also earlier ones, as summarized therein]{2002_coronal_electron_density}. The densities derived around the equatorial plane are expected to be more characteristic of the slow wind, and indeed the density of population \textbf{B} is close the densities derived from coronagraphic observations.   

For the wind populations \textbf{A} to \textbf{C}, the coronal bulk speed is very low below 2 $R_\Sun$ (Figure \ref{fig_med_pts_Helios_PSP_iso_poly}a). This implies large densities (larger than typical coronal densities of about $10^8$ cm$^{-3}$).  However, the iso-poly model is not incorporating several key physical processes of the corona, like thermal conduction and radiative losses, so the region close to the Sun is out of the range modeled by the iso-poly model.  However, in the range $2< r< 25$ $R_\Sun$
the results of \citet{Sheeley1997_coronal_V}, obtained with LASCO coronagraph, are broadly compatible with the velocity profiles of wind populations \textbf{A} to \textbf{D}.

 \begin{figure}[t]
 \hspace{-0.5cm}
 \includegraphics[scale=0.56]{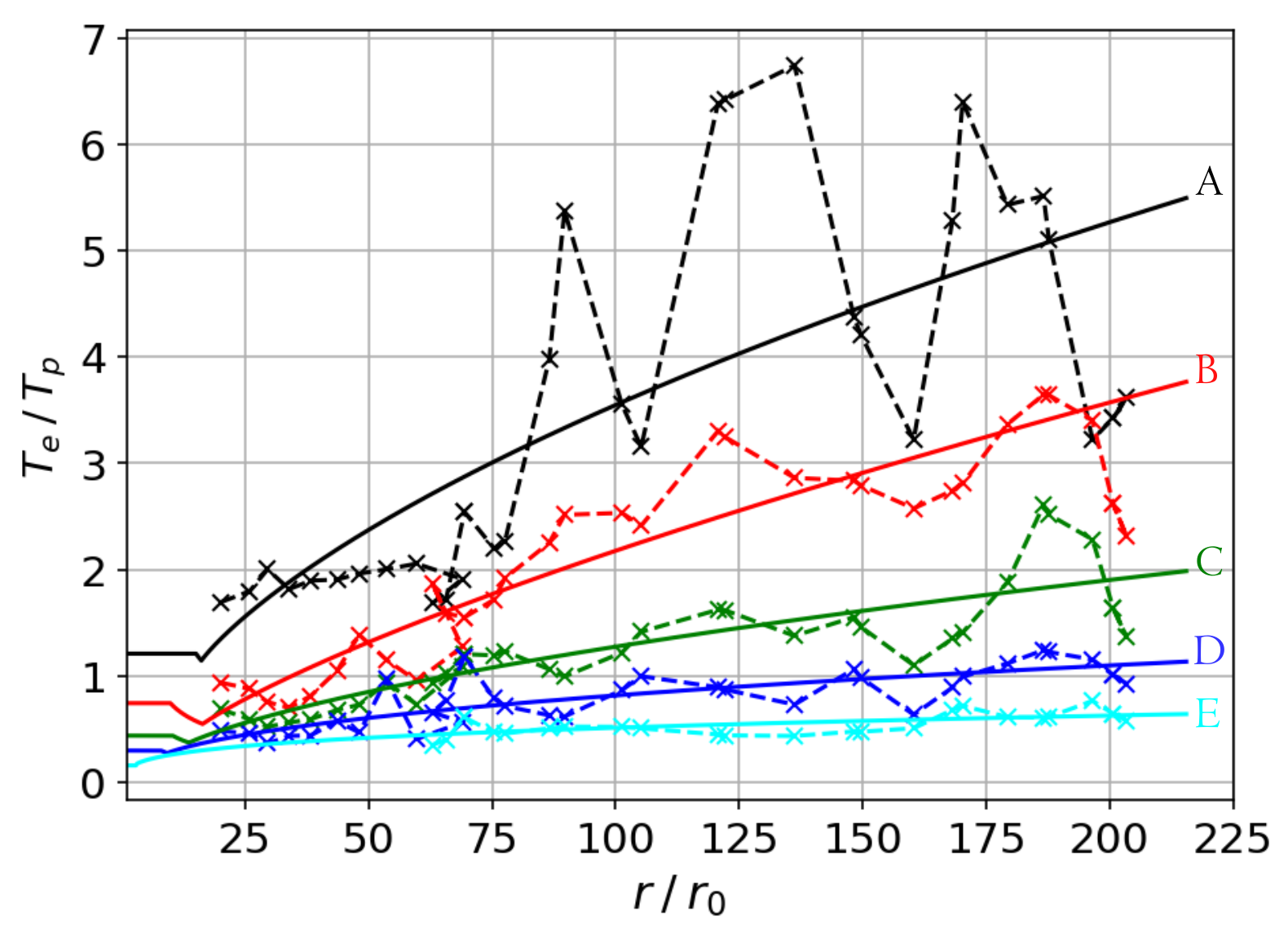}
 \vspace{-0.5cm}
    \caption{Ratio $\Te/\Tp$ for PSP and Helios measurements for all wind populations (dashed lines) as defined in Figure \ref{fig_classif_population_Helios}. The iso-poly numerical solutions, least square fitted to the data (see Figure \ref{fig_med_pts_Helios_PSP_iso_poly}), are shown with solid curves. }
   \label{fig_compar_Pp_Pe}
\end{figure}

The positive correlation bulk speed/proton temperature ($u$,$\Tp$) was originally derived at 1 au \citep{Lopez1986solar}.
The results of the iso-poly model fitted to the in-situ data show that this correlation is kept down to the solar corona (Figure \ref{fig_med_pts_Helios_PSP_iso_poly}a,b). The results of the iso-poly model confirm and extend the results of \cite{Demoulin2009T-Vcorrelation} on the  physical origin of the correlation ($u$,$\Tp$).  This is the result of a dominant wind acceleration by the proton pressure close to the Sun (within $r<20\, R_\Sun$), with a contribution of electron pressure for slower winds.

In contrast, while a clear anti-correlation between the electron temperature and the bulk speed is present above 20 $R_\Sun$, it vanishes closer to the Sun in the iso-poly modeling (Figure \ref{fig_med_pts_Helios_PSP_iso_poly}d).
Indeed, there is no clear trend, and the coronal temperature, $\Teo$, is similar for all wind populations (between 0.7 - 1 MK). Thus, in the iso-poly description, the different wind populations come from solar regions with similar electron temperatures. 

The ratio $\Te/\Tp \: \sim \: \Pe /\Pp $ provides evidence of the species roles in the solar wind dynamics. 
Far from the Sun, $r \: \geq \: 50 \: R_\Sun$, the winds \textbf{A}, \textbf{B} and \textbf{C} are electron driven, the wind \textbf{E} is proton driven, and the wind \textbf{D} has both contributions for $r \: \geq \: 100 \: R_\Sun$ (Figure \ref{fig_compar_Pp_Pe}). 
In the main acceleration region for $r \: < \: 20 \: R_\Sun $, the iso-poly results indicate that the winds are either proton and electron driven (\textbf{A} and \textbf{B}), or proton driven (\textbf{C}, \textbf{D} and \textbf{E}).

The observed electron polytropic indexes, $\game$, are lower than the proton ones, $\gamp$, as shown by \cite{Maksimovic2020anticorrelation} on Helios data, and in Table \ref{tab:param_fit} coupling Helios and PSP data. Then, $\Te$ radially decreases slower than $\Tp$, and they have the possibility to cross each other ($\Te = \Tp $). This is indeed the case for the populations \textbf{B} and \textbf{C}.  It implies that electron pressure is more efficient farther away from the Sun than proton pressure. However, this provides only a weak wind acceleration (Figure \ref{fig_med_pts_Helios_PSP_iso_poly}a).

\section{Conclusion} 
\label{sec:conclusion}

In this paper we have analysed proton and electron solar wind measurements from the instruments SPAN-Ai, SPC and SPAN-E of PSP.
We define five wind populations with the same methodology than the one proposed by \citet{Maksimovic2020anticorrelation} for the Helios data.
We use the overlap distance range of the missions to define the percentage of PSP observations representative of each Helios wind population. 

We find a good agreement between the Helios and PSP wind profiles for the speed, electron density, proton and electron temperature.
The continuous acceleration of the slow solar wind, already shown with Helios data, is also present closer to the Sun in PSP observations.  
Moreover the correlation bulk speed/proton temperature ($u$, $\Tp$) and the anti-correlation bulk speed/electron temperature ($u$, $\Te$), observed at 1 au, are maintained in the PSP observations at least as close as 20 $R_\Sun$ ($\sim$ 0.1 au). 

The polytropic decrease of proton and electron temperatures, previously reported with Helios observations, is extended to the PSP radial range with almost the same polytropic indexes. We have modeled these regions with a fluid approach including separate polytrope behaviours for protons and electrons.  We have no clear evidence that this behaviour changes closer to the Sun with the most recent PSP observations. In order to avoid excessively large coronal temperatures in the model, we impose the polytropic increase of both temperatures to stop at some radial distance, different for electrons and protons. At smaller distances, we simply impose constant temperatures.

The free parameters of the iso-poly model are describing both proton and electron temperature radial profiles. These parameters are determined by a least square fit of the model to the data for $r < 0.5$ au.  This procedure is fully successful to define a model well representing the intermediate solar winds (from 350 - 500 km/s at 1 au). Indeed,  the closeness of the model to the data shows that the observed temperature gradients are sufficient to accelerate such winds with no extra energy required.  

The observations of the slowest wind population show an acceleration over all the observed solar distances.  The iso-poly model, fitted to the data for $r < 0.5$ au, is only able to account for the observed acceleration in this radial range, but not at larger distances.  This result indicates the presence of another source of acceleration which does not heat the plasma and operates on large solar distances, mainly for the slowest solar wind. 

The observed fast wind profiles can be correctly reproduced by the iso-poly model for the Helios data (no PSP data are available for such winds). Nevertheless, the high needed coronal temperature (5 - 6 MK), do not allow to go deeper in the interpretation of the iso-poly modeling results. Indeed, it would require a more complete observational study of the coronal hole temperatures, in order to better estimate the reliability of such modeled coronal temperatures.

We also have found that the electron pressure is dominant, over the proton one, to accelerate the slow winds.  This predominance increases with the solar distance.  For intermediate wind speeds, the proton  pressure is able to provide the main acceleration close to the Sun.  In contrast, the proton pressure is dominant, while not sufficient, to accelerate the fastest wind. 

This paper raises interesting questions about the large distance acceleration processes in the solar wind, as well as about the missing energy to the plasma heating, necessary to describe the observed radial evolution of the slow wind. Indeed, several physical phenomena are candidates to explain a slight acceleration of the solar wind, such as co-rotating interaction regions, Alfven waves, and ambipolar scattering. However, the weight of their respective role in the wind acceleration must be clarified.

\acknowledgments{ Authors thank Parker Solar Probe team for valuable discussions. This work was supported by CNRS Occitanie Ouest, CNES and LESIA. We recognise the collaborative and open nature of knowledge creation and dissemination, under the control of the academic community as expressed by Camille No\^{u}s at http://www.cogitamus.fr/indexen.html.
}

\appendix

\section{Parker Solar Probe Temperature Adjustment between SPAN-Ai and SPC Instruments}
\label{sec:temperature_adjustment_SPAN_SPC}

\begin{figure*}[h]
\hspace{-0.3cm}
\includegraphics[scale=0.56]{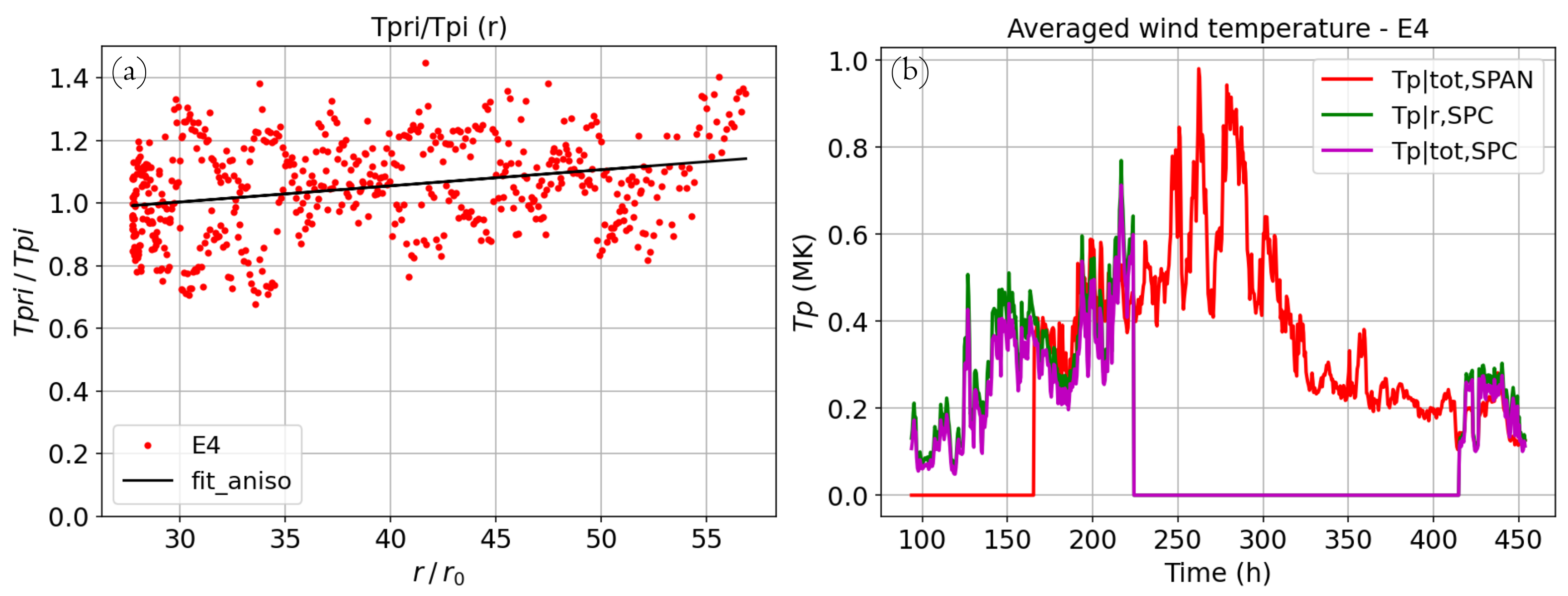}
    \caption{ 
    (a) : SPAN temperature anisotropy $T_{p|r}$/$T_{p|tot}$ (r) over the operating SPAN radial range (red dots), and least square linear fit (solid line).  
    (b) : SPC radial temperature previously adjusted by a factor 2 as mentioned in Section \ref{sec:obs_helios_psp_2} (green), total SPAN temperature (red), and equivalent SPC total temperature using the anisotropy relation between the radial and total temperature from SPAN (purple). 
    }
    \label{fig:appendix_SPC_SPAN_correlation}
\end{figure*}

To have compatible temperatures measured by SPI and SPC covering the largest range of solar distances, we have calibrated SPC temperature with the SPI one using the ratio $T_{r|SPI} / T_{tot|SPI}$ which evolves linearly with solar distance.
The panel (a) of Figure \ref{fig:appendix_SPC_SPAN_correlation} show the distribution of the ratio for the measurements of the encounter 4 (red dots), and a linear adjustment of the form $y = ax + b$ applied to these measurements (solid line). The panel (b) show the SPC temperature adjustment.
The equivalent total proton temperature $T_{tot|SPC}$ (purple curve) is established dividing the radial temperature $T_{r|SPC}$ (green curve) by the anisotropy linear law, assuming the law is extended over the distances covered by SPC. The equivalent total temperature from SPC data extend the SPI one (red curve) on larger distances.
Finally, a unique PSP proton data set is created, associating one time to a unique parameter measurement, so for time where both SPI and SPC measurements are available, the mean value between these two is kept.

\section{Iso-poly Model Detailed Equations}
\label{sec:appendix_iso-poly_detailed_calculs}

In this appendix, we detail the calculation of the iso-poly model. We remind that in the iso-poly description, we limit the $\gams$ to two regions with constant values.
We start from the momentum equation \eqref{eq:momentum_detailed}:
  \begin{align}
    n \, \mp u \frac{d u}{d r} = 
      -  \sums \bigg( \Psisos  \frac{d \ntils^{\gams} }{d r} \bigg) 
      - n \, \mp \frac{G \, M}{r^2}  \,,
    \label{eq:momentum_detailed_appendix}
  \end{align}
where $\ntils = n(r)/\nisos$.  
   
We use the properties of the logarithmic derivative of a composite function to compute:
  \begin{align}
    \frac{d \ntils^{\gams}}{d r} 
    = \ntils^{\gams} \: \frac{d (\gams \ln (\ntils) )}{d r}
    = \gams \ntils^{\gams - 1} \frac{d \ntils}{d r}
    + \ntils^{\gams} \ln (\ntils)  \frac{d \gams}{d r}
  \end{align}
Since we set $\gams$ constant in the isotherm and polytropic regions $d \gams / d r =0 $.
The momentum Equation \eqref{eq:momentum_detailed_appendix} is rewritten as:
  \begin{align}
    n \, u \frac{d u}{d r} = 
     -  \sums \bigg( 
       \frac{\Psisos}{\mp} \gams \, \ntils^{\gams - 1} \frac{d \ntils}{d r}
           \bigg) 
    - n \frac{G \, M}{r^2}  
  \end{align}
To further simplify the equation writing, we define $\cs$ for the species s 
as an equivalent sound speed: 
  \begin{align}
    \cs^2 = \frac{\gams \Psisos}{ \mp \nisos }
          = \frac{\gams \kB \Tsisos}{\mp}
    \label{eq:appendix_cs}
  \end{align}
We also define the variable $\xs$ with $\xs = \ntils^{\gams - 1}$. With the above definitions, we obtain: 
  \begin{align}
    n \, u \frac{d u}{d r} = 
    -  \sums \nisos \bigg(  \cs^2 \xs \frac{d \ntils}{d r} 
\bigg) 
    - n \frac{G \, M}{r^2}  
    \label{eq:appendix_As_Bs}
\end{align}

Next, the conservation of mass flux writes as $n(r) = \Cn / (u\, r^2)$, where $\Cn$ a constant. 
Developing the calculation of the derivative of $\ntils(r)$: 
  \begin{align}
   \frac{d \ntils}{d r} = - \frac{1}{\nisos} \frac{\Cn}{r^2} \bigg[ \frac{2}{ur} + \frac{1}{u^2} \frac{du}{dr} \bigg]
  \end{align}
Finally, the momentum equation is written as: 
  \begin{align}
   & \frac{du}{dr} 
     \bigg[ \frac{\Cn}{u r^2} u - \sums \frac{\Cn}{r^2} \frac{\cs^2}{u^2 } \xs \bigg]
     = \sums \bigg[ \frac{\Cn}{r^2} \frac{2}{u r} \cs^2 \xs  \bigg]
     - \frac{\Cn}{u r^2} \frac{G \, M}{r^2} \\
\Rightarrow \quad
   & \frac{du}{dr} 
     \underbrace{ \bigg[ 1 - \sums \frac{\cs^2}{u^2} \xs \bigg] }_{a(r,u)} 
     = \underbrace{  \frac{1}{u r} \bigg[ \sums  2 \cs^2 \xs - \frac{G \, M}{r} \bigg] }_{b(r,u)} 
\label{eq:appendix_momentum_all_term_detailled}
\end{align}

The equation \eqref{eq:appendix_momentum_all_term_detailled} summarizes the iso-poly model. 
The solar wind flow is described by the transonic solution with $du/dr \neq 0$ for all $r$ values. Then, where $b(r,u)=0$, $a(r,u)$ should also vanishes. This defines the so called critical or sonic point.   If it is located in the isothermal region $\xs=1$, and the critical point is defined by the isothermal Parker's result:

  \begin{align}
    \uc = \sqrt{ \cp^2 + \ce^2 } 
        \quad \text{and} \quad
    \rc = \frac{G \, M}{ 2 ( \cp^2 + \ce^2 )}
  \label{eq:appendix_uc_rc}
  \end{align}
with $\cp$ and $\ce$ computed with Equation \eqref{eq:appendix_cs} and $\gams=1$.
In the polytropic region, the critical radius $\rc$ is divided by the factor $\gams >1$ compared to Equation \eqref{eq:appendix_uc_rc}.   When we fit the iso-poly model to observations (Section \ref{sec:obs_model_3}), the optimum $\rc$ value stays within the isothermal region. Moreover, if during the fitting iteration $\rc$ goes a bit in the polytropic region, its value is divided by $\gams$, which bring it back to the isothermal region. The temperatures $\Tp(\rc)$ and $\Te(\rc)$ would need to be significantly lower than $\Tso$ to keep $\rc$ in the polytropic region.

The optimal way to compute the transonic solution is to use an asymptotic development around $\rc$ of the equation \eqref{eq:appendix_momentum_all_term_detailled} to get the slope $du/dr$ at the critical point.  In fact,  we proceed simpler using a tiny positive (resp. negative) shift from ($\rc,\uc$) to integrate upward (resp. downward).   With a tiny shift such solutions converge rapidly toward the transonic solution, thanks to the hyperbolic topology present around the critical point. Finally, with $u(r)$ computed, the density expression is deduced from mass flux conservation, and the temperature radial profile of the species is defined by Equation \eqref{eq:expression_Ts}.

\section{Determination of Iso-poly Parameters with the $\chi^2$ Test}
\label{sec:appendix_identif_param}

\begin{figure*}[h]
\hspace{-0.3cm}
\includegraphics[scale=0.59]{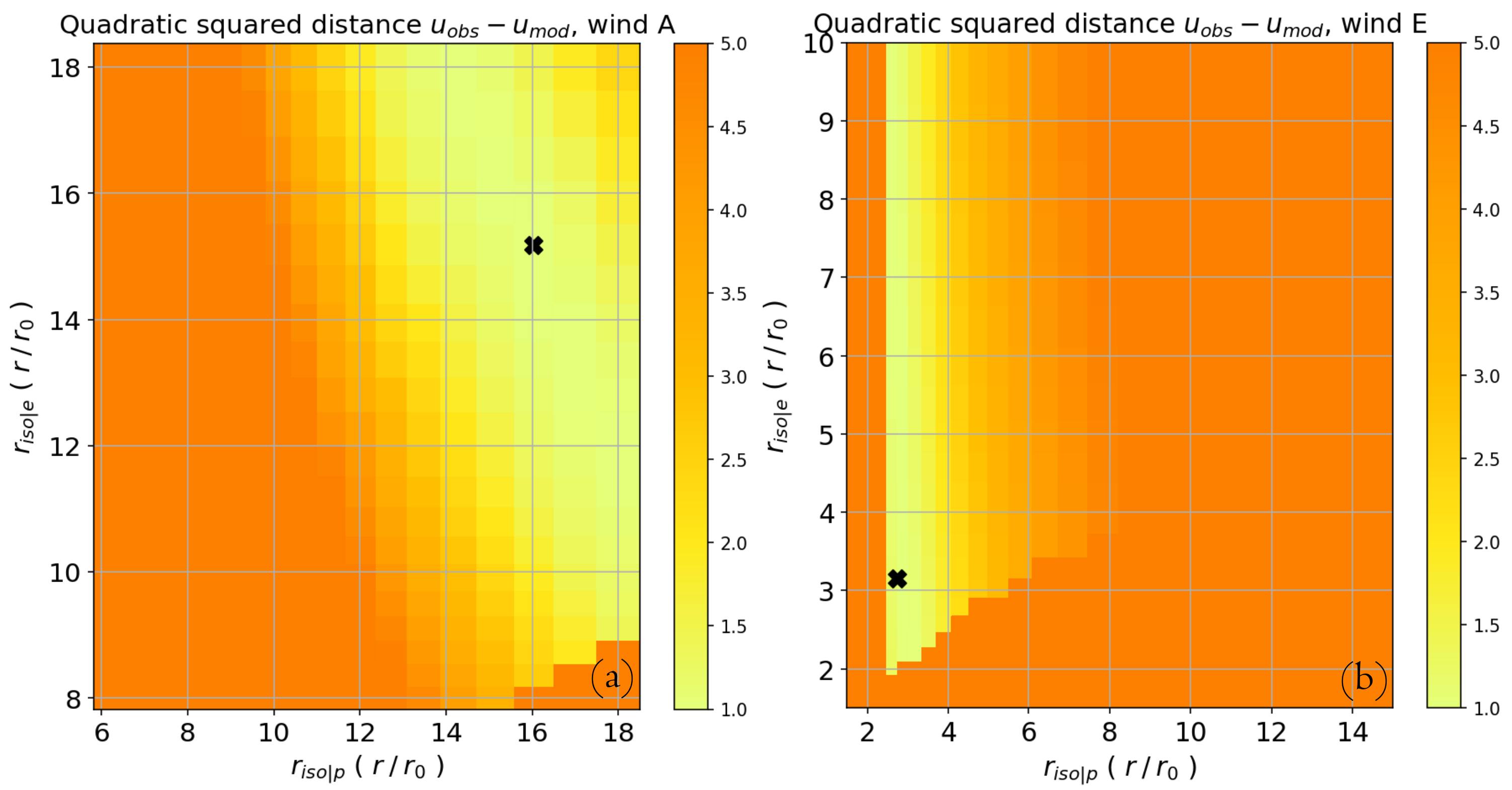}
    \caption{ 
    $\chi^2$ values expressing the distance between the iso-poly model and the solar wind data within the ($\risop$, $\risoe$) plane.  (a) the slowest population A (Helios-PSP data), (b) the fastest population E (Helios data).
    The black crosses represents the location of the best set of parameters (main minimum of $\chi^2(\risop, \risoe)$).
    }
    \label{fig:appendix_chi2_map}
\end{figure*}

We outline below the determination of the free parameters for the iso-poly model. First the modeled temperature profile are least square fitted to each observed profile.  This defines the polytropic indexes $\gamp$ and $\game$. To be in accordance with the in-situ measured temperature, we constrain for each population of the model, the minimal coronal temperature $\Tso$ to the closest radially observed temperature. Next, the model bulk speed is compared to observed velocities for different ($\risop$, $\risoe$) values, and for each wind population. The resulting $\chi^2$ minimization map ($\risop$, $\risoe$) for the population \textbf{A} and \textbf{E} are plotted on Figure \ref{fig:appendix_chi2_map}. 
With the supposed continuity of the temperature profiles, this also determines the coronal temperatures $\Tpo$ and $\Teo$ (supposed to be uniform below $ \risop$ and $\risoe$, respectively).

The optimal set of parameters is not located in a spot minimum region of the map, but in a valley. For the wind \textbf{A} both $\risoe$ and $\risop$ values have an influence, because the valley is diagonally oriented. The smaller $\risop$ is, the bigger $\risoe$ is, so that they compensate each other to fit as best as possible to the observed speed profile. Indeed, Figure \ref{fig_compar_Pp_Pe} shows that $\Tp \approx \Te$, so $\Pp \approx \Pe$ in the main acceleration region ($r < 20 R_\Sun$).  In contrast, the wind \textbf{E} minimization maps shows a region of smaller values on a more vertically oriented valley, signifying that the value of $\risop$ is well determined, and thus is more determinant in the modeling of fast wind than $\risoe$. Indeed, Figure \ref{fig_compar_Pp_Pe} shows that $\Tp$, then $\Pp$, is dominant for wind \textbf{E} in the main acceleration region.


\end{document}